# SPIN-WAVE NONRECIPROCITY BASED ON INTERBAND MAGNONIC TRANSITIONS

K. Di, H. S. Lim[*], V. L. Zhang, S. C. Ng, and M. H. Kuok[†]

*Department of Physics, National University of Singapore, Singapore 117542, Singapore*

[*]e-mail: phylimhs@nus.edu.sg; [†]e-mail: phykmh@nus.edu.sg

**ABSTRACT**

We theoretically demonstrate linear spin-wave nonreciprocity in a $Ni_{80}Fe_{20}$ nanostripe waveguide, based on interband magnonic transitions induced by a time-reversal and spatial-inversion symmetry breaking magnetic field. An analytical coupled-mode theory of spin waves, developed to describe the transitions which are accompanied by simultaneous frequency and wavevector shifts of the coupled spin waves, is well corroborated by numerical simulations. Our findings could pave the way for the realization of spin-wave isolation and the dynamic control of spin-wave propagation in nanoscale magnonic integrated circuits via an applied magnetic field.

Besides being of scientific interest, the phenomenon of the nonreciprocal propagation of waves is of great technological importance. From the perspective of applications, nonreciprocity provides an extra degree of control in molding the flow of waves in fields ranging from mature microwave technology[1] to the rapidly expanding areas of photonics,[2-4] magnonics[5-8] and phononics.[9,10] Furthermore, it is crucial to the stabilization of integrated circuits and the suppression of undesirable crosstalk and interference arising from imperfection-induced scattering.[3]

While much work has been undertaken on optical nonreciprocity, relatively little has been done on nonreciprocity involving spin-wave propagation. Research into spin waves (SWs) in the area of magnonics[11-20] of nanomagnets, is rapidly gaining interest due to attributes like their short wavelengths and highly tunable dynamic properties.[13] Because of their low energy dissipation, SWs in magnetic insulators[12,16-18] find applications in spintronics as an ideal spin current carrier. In these research fields, nonreciprocity of SWs is an important property for various functionalities. Thus far, studies of magnonic nonreciprocity[5-8] focused on the long-wavelength surface Damon-Eschbach (DE) waves with propagation direction perpendicular to the applied magnetic field. Such a nonreciprocity originates from the lower symmetry at sample surfaces and dipole-dipole interactions rather than from isotropic exchange interactions.[21] However, in the sub-100nm range, SWs are exchange-dominated and nonreciprocity due to dipole-dipole interaction is less significant. Hence, nanoscale miniaturization of magnonic devices necessitates the realization of nonreciprocity of exchange-dominated SWs.

In this Letter, we demonstrate spin-wave nonreciprocity arising from a new mechanism, namely, interband magnonic transitions. We employ micromagnetic simulations and our analytical coupled-mode theory of exchange SWs to show that nanoscale linear nonreciprocity of exchange-dominated SWs in a nanostripe waveguide can be achieved by applying a time-reversal and space-inversion symmetry-breaking magnetic field. Depending on propagation direction, SWs propagating in the waveguide could undergo interband magnonic transitions. It is to be noted that the previously studied nonreciprocity is completely different from our proposed nonreciprocity. First, the former originates from the dipole-dipole interaction, while ours arises from interband magnonic transitions and is present even in the absence of the dipole-dipole interaction. Second, while frequency and wavevector are reciprocal quantities in uniform thin films for the former case,[21] they are no longer so in our case. Our proposed nonreciprocity could lead to nonreciprocal magnonic devices, such as spin-wave isolators which serve as building blocks in integrated magnonics.

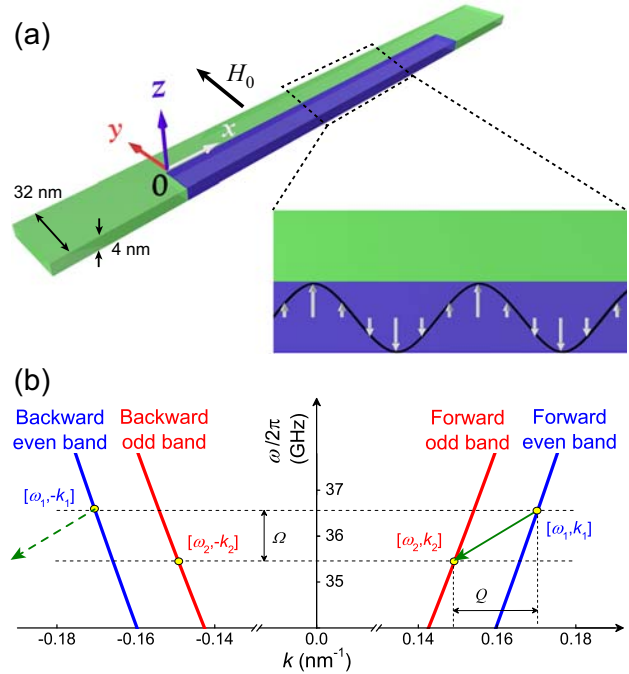

FIG. 1. Nonreciprocal spin-wave mode transitions in the Permalloy nanostripe waveguide. (a) The perturbation magnetic field is applied in the *y*-direction within the blue region of the waveguide. The enlarged view shows the instantaneous profile of the perturbation field vector represented by arrows. (b) Dispersion relations of spin waves with blue and red lines representing the respective even and odd symmetry bands. The solid (dashed) green arrow indicates shifts of frequency and wavevector in the forward (backward) propagation direction due to the perturbation field. Resonant coupling exists between modes $[\omega_1, k_1]$ and $[\omega_2, k_2]$ in the forward (+x) direction, while coupling between modes $[\omega_1, -k_1]$ and $[\omega_2, -k_2]$ in the backward (–x) direction is insignificant.

The waveguide investigated is an 8μm-long $Ni_{80}Fe_{20}$ (Permalloy) nanostripe of rectangular cross section, with a width $w$ of 32 nm and a thickness of 4 nm, as depicted in Fig. 1(a). A bias external magnetic field $H_0 = 0.3\,T/\mu_0$, where $\mu_0$ is the vacuum permeability, is applied in the +y-direction. Figure 1(b) shows the lowest and second-lowest energy bands, with respective even and odd symmetry, of the magnonic dispersion relations of the waveguide calculated from micromagnetic simulations.[22] For brevity, a spin-wave mode with angular frequency $\omega_n$ and wavenumber $k_n$ will be labeled $[\omega_n, k_n]$, where $n = 1, 2$. For the linearized Landau-Lifshitz-Gilbert (LLG) equation, the dynamic magnetization of SWs[23,24] is given by

$$\boldsymbol{m}_n = \begin{pmatrix} m_{x,n} \\ m_{z,n} \end{pmatrix} = \begin{pmatrix} 1 \\ i \end{pmatrix} f_n(y) exp[i(k_n x - \omega_n t)], \qquad (1)$$

where the mode profile $f_n(y) = a_n \cos[\kappa_n(y+w/2)]$, and $k_n$ is the wavenumber along the x-axis. In the definition of $f_n$, the effective transverse wavevector $\kappa_n = p_n \pi / w$, where $p_n = 0, 1, 2, \ldots$ is the quantization number of mode $\boldsymbol{m}_n$ (free-spin boundary condition[24]), $a_n$ the normalization constant determined by $\int_{-w/2}^{w/2} f_n^2 dy = w$, and $w$ the width of the waveguide. Note that $f_n$ constitutes a complete orthogonal basis[23] over the interval $-w/2 \leq y \leq w/2$, and that $f_n = 1$ and $f_n = \sqrt{2}\cos[\pi/w(y+w/2)]$ for the lowest and second-lowest energy bands respectively.

In analogy to indirect electronic transitions in semiconductors and indirect photonic transitions in photonic structures[2,4] (indirect transitions involve simultaneous shifts in energy and momentum), indirect transitions between two spin-wave modes within the waveguide can be induced by a perturbation magnetic field $h'$, of frequency $\Omega$ and wavevector $Q$, applied along the $y$-direction

$$h'(x,y,t) = h_0(y)\cos(Q x - \Omega t), \qquad (2)$$

where the amplitude of the field is defined by $h_0(y) = 0.05$ T/$\mu_0$ for $y < 0$, and $h_0(y) = 0$ T/$\mu_0$ for $y > 0$ [see Fig. 1(a)]. Significant transition between two spin-wave modes will occur only if the phase-matching conditions,[25] namely $Q = k_2 - k_1$ and $\Omega = \omega_2 - \omega_1$, where 1 and 2 denote the respective initial and final modes, are approximately satisfied.

The analytical coupled-mode theory explaining the exchange SWs mode transition dynamics is developed as follows. As the considered spin-wave wavelengths (around 30 nm) lie within the sub-100nm range where the exchange interaction dominates,[14] it is a reasonable approximation to disregard dipole-dipole interactions in the analysis. Therefore, neglecting the Gilbert damping term, the linearized LLG equation for the Permalloy waveguide under field $h'$ can be written as

$$\frac{1}{\gamma}\frac{d}{dt}\boldsymbol{m} = \left(\hat{H}_{ex} + \hat{H}_0 + \hat{H}'\right)\boldsymbol{m}, \qquad (3)$$

where the dynamic magnetization $\boldsymbol{m} = \begin{pmatrix} m_x \\ m_z \end{pmatrix}$, $\hat{H}_{ex} = M_S D(\partial^2/\partial x^2 + \partial^2/\partial y^2)\hat{\sigma}$, $\hat{H}_0 = -H_0\hat{\sigma}$, $\hat{H}' = -h'\hat{\sigma}$, $\hat{\sigma} = \begin{pmatrix} 0 & -1 \\ 1 & 0 \end{pmatrix}$, and $D = 2A/(\mu_0 M_S^2)$. In the following calculations, the saturation magnetization $M_S$, exchange constant $A$ and gyromagnetic ratio $\gamma$ of Permalloy, were set to $8\times 10^5$ A/m, $1\times 10^{-11}$ J/m and $2.211\times 10^5$ Hz·m/A, respectively.

We assume that the perturbation field $h'$ approximately satisfies the phase-matching conditions $k_2 - k_1 - Q = \Delta k \approx 0$, and $\omega_2 - \omega_1 - \Omega = 0$, where $k_2 = 0.149$ nm$^{-1}$, $k_1 = 0.170$ nm$^{-1}$, $\omega_2/2\pi = 35.5$ GHz, and $\omega_1/2\pi = 36.6$ GHz. In the presence of the weak perturbation field $h'$ (i.e. $\hat{H}' \ll \hat{H}_{ex} + \hat{H}_0$), which propagates in the +x direction, $\boldsymbol{m}_n$ are no longer eigenmodes, and the SWs is expressible as a linear combination of the basis $\boldsymbol{m}_n$, i.e.,

$$\boldsymbol{m} = \sum_n C_n(x)\boldsymbol{m}_n, \tag{4}$$

where $n$ spans all the eigenmodes, and $C_n$ are coefficients that reflect the mode coupling along the waveguide. Using Eqs. (1) to (4) and the slowly varying amplitude approximation[25] for $C_n$ in which terms containing $d^2C_n(x)/dx^2$ are neglected, one obtains

$$\sum_n 2ik_n M_S D \frac{dC_n(x)}{dx}\boldsymbol{m}_n - \sum_n h'C_n(x)\boldsymbol{m}_n = 0. \tag{5}$$

Taking the inner product of the above equation with $\boldsymbol{m}_1^*$ the complex conjugate of $\boldsymbol{m}_1$, and integrating over $y$, yields

$$4ik_1 M_S Dw \frac{dC_1(x)}{dx} = \sum_n C_n(x)\left(e^{i\delta k_{n+}x}e^{i\delta\omega_{n+}t} + e^{i\delta k_{n-}x}e^{i\delta\omega_{n-}t}\right)\int_{-w/2}^{w/2} h_0(y)f_1 f_n dy, \tag{6}$$

where $\delta k_{n\pm} = k_n - k_1 \pm Q$, and $\delta\omega_{n\pm} = \omega_n - \omega_1 \pm \Omega$.

Although Eq. (6) contains an infinite number of terms on the right, imposition of the phase-matching conditions reduces the number to just one. First, due to energy conservation, the occurrence of mode transition necessitates exact fulfillment of the condition[2] $\omega_n = \omega_1 \pm \Omega$. Second, we need only retain the phase-matching terms with $\delta k_{n\pm} \approx 0$ on the right-hand side, and neglect the phase-mismatch terms, because the integral of the latter, over sufficiently long distance in the $x$-direction, vanishes,[25] indicating zero contribution. The phase matching conditions reduce Eq. (6) to a two-state problem, namely,

$$\frac{d}{dx}\begin{pmatrix} C_1(x) \\ C_2(x) \end{pmatrix} = \begin{pmatrix} 0 & -ie^{-i\Delta k x}K/J_1 \\ -ie^{i\Delta k x}K/J_2 & 0 \end{pmatrix}\begin{pmatrix} C_1(x) \\ C_2(x) \end{pmatrix}, \tag{7}$$

where the overlap integral $K = \int_{-w/2}^{w/2} h_0(y)f_1 f_2 dy$ and $J_n = 4k_n M_S Dw$ $(n=1, 2)$. It should be emphasized that mode transition is possible for any two modes irrespective of their symmetry, provided that the phase-matching conditions are satisfied, and the coupling coefficient $K$ has a nonzero value. It is to be noted that to fulfill the latter condition, the perturbation field can be of a form different from that of Eq. (2), which is chosen for simplicity of calculations.

Given the initial conditions $C_1(0)=1$ and $C_2(0)=0$, the solutions to Eq. (7) are

$$C_1(x) = e^{-ix\Delta k/2} \left[ \cos\left(x\sqrt{a+(\Delta k/2)^2}\right) + i \frac{\Delta k/2}{\sqrt{a+(\Delta k)^2}} \sin\left(x\sqrt{a+(\Delta k/2)^2}\right) \right]$$

$$C_2(x) = \frac{-i e^{ix\Delta k/2}}{\sqrt{1+\Delta k^2/4a}} \sqrt{\frac{J_1}{J_2}} \sin\left(x\sqrt{a+(\Delta k/2)^2}\right),$$
(8)

where $a = K^2/(J_1 J_2)$. If the phase-matching conditions are exactly satisfied (i.e. $\Delta k = 0$), the mode transition will be complete at integral multiples of the coupling length $l_c = \pi/(2\sqrt{a})$, defined as the propagation distance over which one mode undergoes a complete transition to the other. Note that $l_c$ is inversely proportional to the strength of the field $h_0$. If $\Delta k \neq 0$, then $|C_1|$ never vanishes, indicating incompletion of the mode transition and energy transfer. However, for strongly phase-mismatching cases ($\Delta k \gg \sqrt{a}$), e.g., when SWs propagate in the reverse (-x) direction, we have $|C_2| \ll |C_1|$, indicating that no significant mode transition will occur. Hence, only for certain pairs of $Q$ and $\Omega$ values will the phase-matching conditions be simultaneously satisfied for transitions between modes [$\omega_1$, $k_1$] and [$\omega_2$, $k_2$], but not for those between mode [$\omega_1$, $-k_1$] and any other mode [see Fig. 1(b)]. This means that mode transitions will occur only for one propagation direction, indicating that spin-wave propagation in the waveguide is nonreciprocal. Of note is that this nonreciprocity phenomenon is due to the breaking of time-reversal and spatial-inversion symmetry induced by the perturbation magnetic field.

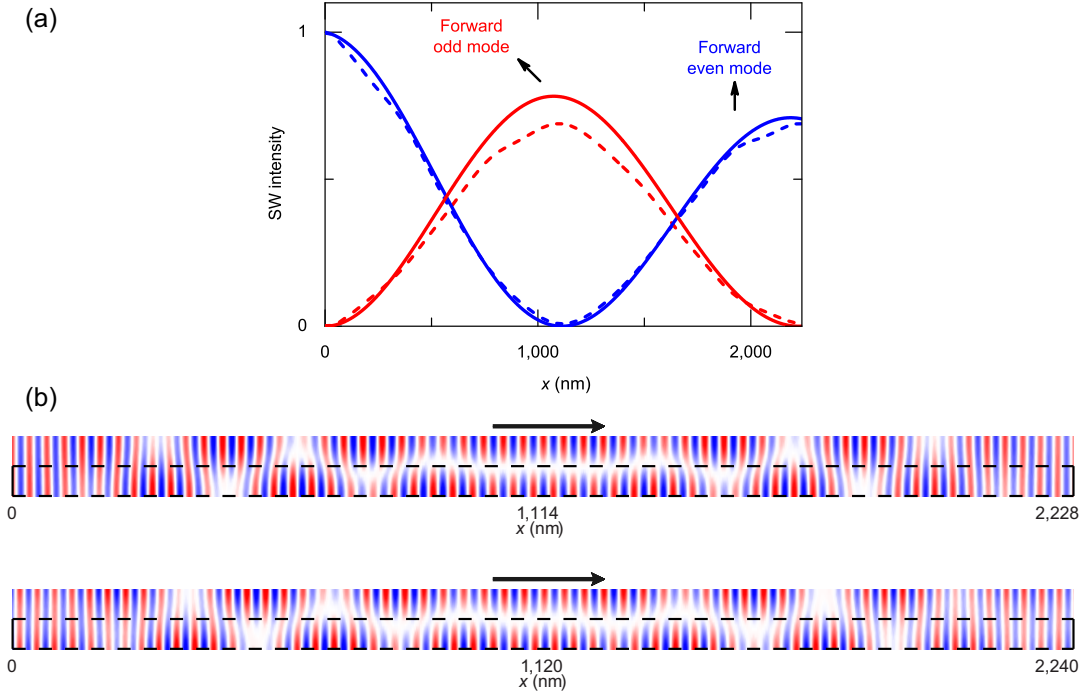

FIG. 2. Forward spin-wave mode transitions in the waveguide. (a) Intensities of coupled magnon modes $[\omega_1, k_1]$ (blue curves) and $[\omega_2, k_2]$ (red curves) obtained from the coupled-mode model (solid curves) and micromagnetic simulations (dashed curves), respectively. (b) *x-y* view of instantaneous dynamic magnetization $m_z$ based on the coupled-mode theory (top panel) and micromagnetic simulations (bottom panel). Dashed boxes denote the regions where the perturbation field is applied. Transitions between the two modes result in the mode profiles (across the waveguide width) to gradually evolve, with increasing *x*, from being even to odd and back to even symmetry. The coupling lengths are determined to be 1114 and 1120 nm based on the coupled-mode theory and the micromagnetic simulations, respectively. The magnitudes of $m_z$ are color-coded, with red, white and blue representing positive maximum, zero and negative maximum, respectively. The arrows indicate the propagation of the two modes in the forward (+x) direction.

The coupling length was calculated to be 1114 nm. The intensities of modes $[\omega_1, k_1]$ and $[\omega_2, k_2]$, evaluated from Eq. (8), are presented in Fig. 2(a). Due to Gilbert damping, the intensities of modes along the $x$-direction are modified by a factor $e^{-x/\lambda}$, where the spin-wave decay length[19] $\lambda = v_g/[2\alpha\omega]$, $v_g$ being the spin-wave group velocity and $\alpha$ the Gilbert damping coefficient. The corresponding instantaneous profile of the $z$-component $m_z$ of the dynamic magnetization is presented in Fig. 2(b).

Micromagnetic simulations were performed to investigate the dynamics of the predicted spin-wave nonreciprocity by solving the LLG equation using the OOMMF package[26] based on a unit cell size of $2\times2\times4$ nm$^3$ and $\alpha = 5\times10^{-4}$, with both dipolar and exchange interactions taken into account and neglecting magnetocrystalline anisotropy. Unwanted SW reflections were eliminated by setting the Gilbert damping to 1.0 at both ends of the waveguide.

To determine the coupling length $l_c$, the magnetic-field perturbation region was first chosen to be sufficiently long along the $x$-direction, i.e., for the occurrence of two complete mode transitions. A continuous SW of mode $[\omega_1, k_1]$ was excited at the left end of the waveguide as the initial wave. The simulated intensities $|C_1|^2$ and $|C_2|^2$ of the respective modes $[\omega_1, k_1]$ and $[\omega_2, k_2]$ as a function of interaction distance within the perturbation region, obtained by performing a fast Fourier transform of the out-of-plane component $m_z$ of the dynamic magnetization in time, are displayed in Fig. 2(a). The simulated coupling length is about 1120 nm. The simulated data accord well with the coupled-mode theory in terms of both the relative mode intensities and the coupling length. Both methods indicate that the energy is transferred back and forth between the coupled modes. It it noteworthy that the simulated coupling length is found to be inversely proportional to the strength of the perturbation field, consistent with the coupled-mode theory. Figure 2(b) shows that the simulated instantaneous dynamic magnetization reproduces well the theoretically calculated one. It also reveals that the forward-propagating mode undergoes a transition, with the symmetry of its $m_z$ profile progressively evolving from even to odd and finally reverting to even.

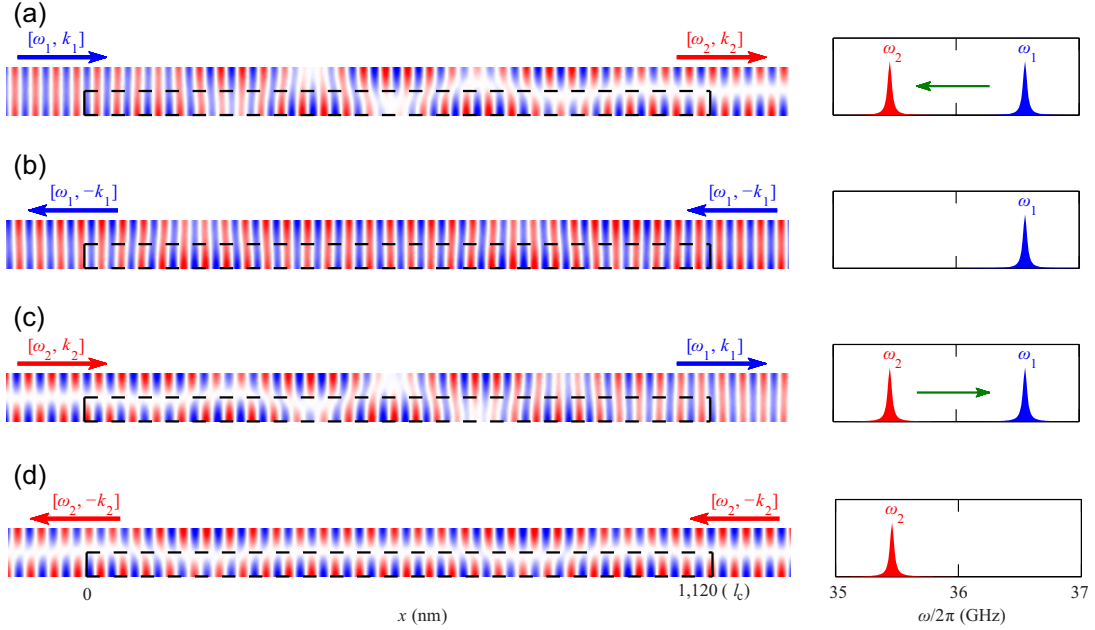

FIG. 3. Snapshots of out-of-plane components $m_z$ of spin waves. (a) Spin-wave $[\omega_1, k_1]$ propagating forward (from left to right) undergoing a complete transition to mode $[\omega_2, k_2]$. (b) Spin-wave $[\omega_1, -k_1]$ propagating backward (from right to left) remaining unaltered. (c) Spin-wave $[\omega_2, k_2]$ propagating forward undergoing a complete transition to mode $[\omega_1, k_1]$. (d) Spin-wave $[\omega_2, -k_2]$ propagating backward remaining unaltered. The corresponding mode frequency changes are indicated by panels on the right. Dashed boxes denote regions where the perturbation field is applied.

To demonstrate the concept of nonreciprocal mode transitions, the length of the perturbation region along the $x$-axis was then set to the simulated coupling length $l_c$. The frequencies and propagation directions of the initial wave were varied by changing the frequency and location (left or right end of the waveguide) of the excitation field. In order to excite SWs with the desired (even or odd) symmetry, the applied excitation field should possess the same symmetry, namely, even or odd, respectively, as established by Di *et al.*[22] The snapshot of $m_z$, presented in Fig. 3(a), reveals that the symmetry of the mode profile of the right-propagating SW is completely converted from even to odd. The right panel of Fig. 3(a) shows that the corresponding frequency spectrum changes from $\omega_1$ to $\omega_2$ after passing through the perturbation region. In contrast, on reversal of the propagation direction, i.e., when SWs of mode $[\omega_1, -k_1]$ were considered, no changes in their mode profiles or frequencies were found [see Fig. 3(b)], indicating the occurrence of negligible mode transition. Thus, the waveguide exhibits time-reversal symmetry breaking, resulting in drastic contrasting effects for the two opposite propagation directions. The same nonreciprocal effects were also observed for the initial modes $[\omega_2, k_2]$ and $[\omega_2, -k_2]$ as illustrated in Figs. 3(c), (d).

The magnonic nonreciprocity presented here has the following merits. First, being valid for exchange-dominated SWs, it could in principle permit the realization of nanoscale integrated magnonic circuits. Second, the linearity of the phenomenon makes its realization more straightforward and energy-efficient than that involving nonlinear effects. Recently, it was reported that ultrafast dynamic control of photonic[27,28] and magnonic[11,13] structures offers linear approaches for achieving novel functionalities, such as the frequency shift of waves, which usually involves nonlinear effects in static artificial materials, i.e. those with time-invariant structural arrangements. We have shown, from another perspective, that the dynamic modulation of a magnonic waveguide can give rise to hitherto unreported nonreciprocal propagation of SWs.

In conclusion, we have developed a coupled-mode model for exchange SWs for analyzing spin-wave dynamics under a perturbation magnetic field. Employing this model, we demonstrated that application of the field can effect nonreciprocal spin-wave interband transitions arising from the simultaneous destruction of time-reversal and spatial-inversion symmetry. Micromagnetic simulations yielded results which accord well with those of the coupled-mode theory. The small discrepancy observed is partially ascribed to the neglect of dipolar interactions in the theory. Importantly, this spin-wave nonreciprocity could be exploited for the design of nonreciprocal magnonic nanoscale devices, which could serve as building blocks of integrated magnonic and spintronics circuits. For instance, when used in conjunction with spin-wave filters,[29] our proposed nonreciprocal waveguide could function as a SW isolator.

Financial support by the Ministry of Education, Singapore under Grant No. R144-000-282-112 is gratefully acknowledged.